# CONTRIBUTIONS TO INTERFRAME CODING

Marcos Faúndez-Zanuy, Francesc Vallverdú-Bayes, Francesc Tarrés-Ruiz.

Signal Theory & Communications Department. ETSETB. UPC.

C/ Gran Capità ,s/n mòdul D5

08034 BARCELONA (SPAIN)

tel.:93-4016440

fax:93-4016447

e-mail:marcos@tsc.upc.es

## ABSTRACT

Advanced motion models (4 or 6 parameters) are needed for a good representation of the motion experimented by the different objects contained in a sequence of images. If the image is split in very small blocks, then an accurate description of complex movements can be achieved with only 2 parameters. This alternative implies a large set of vectors per image.

We propose a new approach to reduce the number of vectors, using different block sizes as a function of the local characteristics of the image, without increasing the error accepted with the smallest blocks.

A second algorithm is proposed for an inter/intraframe coder.

## I. INTRODUCTION

This paper describes complementary techniques for interframe coding, with the goal of reducing the total amount of information.

In many real scenes, the motion vector information is the same in big areas of the image, thus an efficient coding implies a variable block size. One possibility, is to adopt a variable block shape, segmenting the image in homogeneous parts, which has several problems:
1. Often it is difficult to segment the image accurately.
2. Segmentation implies large computational complexity.
3. The block shape coding requires a lot of bits and often, the transmission ratio is increased.

Another solution [3] implies the calculation of motion vectors with small and big size, and the selection of the big size when the error is similar to the accumulated error of the equivalent small blocks.

We propose the use of variable block size with a predetermined shape. The minimum and maximum block size is also fixed and studied in our paper.

## II. PARAMETERS SELECTION

In our study, we have selected the following parameters:
* error criterion: mean absolute frame difference (MAD).
* motion model: two parameters ($d_x$,$d_y$), although this method is easily extensible to more complicated models.

___________________________________________________________________

* motion estimation algorithm: modification of the conjugate direction algorithm [4], with maximum displacement allowed fixed to 7.
* prediction process: writing method proposed in [4]. The motion vectors show the address in the image $I_k$ where must be written the block of the image $I_{k-1}$.
* lower block size: we have studied the cases 8x8 and 16x16.

The justification of the different parameters selected can be found in [4].

III. VECTOR POSTPROCESSING

The motion vectors obtained with the motion estimation algorithm, are coded with a quadtree structure (fig. 1) which is used to provide some important advantages:

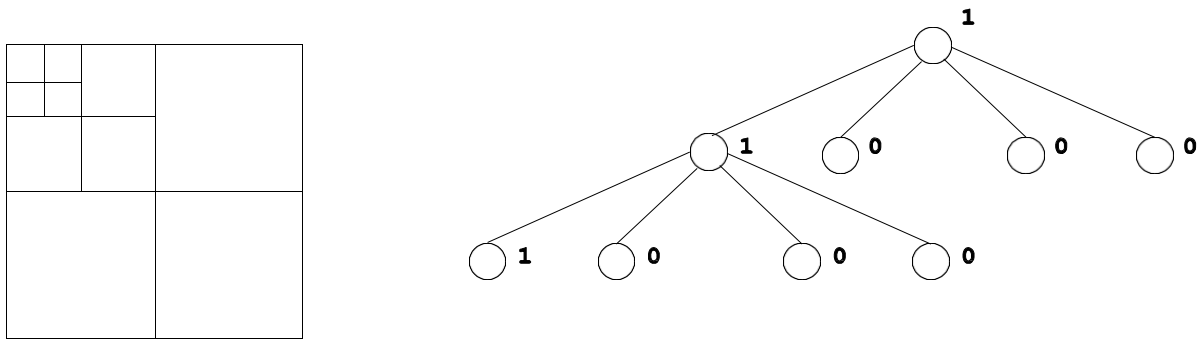

fig. 1 quadtree and codification for sizes from 8x8 to 64x64.

1. It is very simple to encode and decode.
2. It requires a few bits on its coding.
3. The compression ratio is increased about 3 times for a videoconferencing scenes.

This structure has been successfully used in different references ([1], [2]) and applications.

The quadtree structure is used to efficiently encode the motion information obtained with a modification of the conjugate direction method proposed in [4].

III.1 TREE CONSTRUCTION.

The block size selection is parallel to the tree construction. There are two ways to construct the tree:
a) top-down: from the big size, we split it into four squares if the vectors contained are not identical. Every square is treated in the same way until the minimum size is reached or the square is considered homogeneous.
b) bottom-up: from the smallest size, four adjacent blocks are grouped if the four vectors are identical. The process is repeated until the maximum size is reached or it is not possible the clustering.

We have selected the second alternative. If the clustering is realized only when the four vectors are identical, both methods obtain the same result.

Computational time, would be different. If the image contains big areas with identical vectors, the first alternative would be selected. If not, the second alternative is more suitable.

The compression ratio can be increased if the condition of equality between vectors is relaxed to "nearly equal" vectors. This alternative will increase the error in the postprocessing stage, that otherwise is not affected. Sometimes, it will reduce the error, because it is similar to a smoothing function of the vectors, that can remove erroneous motion estimations.

IV. INTER/INTRAFRAME CODING.

For an inter/intraframe predictor, we propose the following structure in order to encode jointly the block shape and the inter/intraframe decision:

1. To implement the quadtree coding proposed in the last sections, grouping blocks coded with the same predictor.
2. Last level of the tree is used to indicate what kind of predictor has been used (inter or intra). It is interesting to observe that this last level was not coded in the original quadtree structure, because it indicates no division, and this is already known from the number of levels used. Also, in the terminal nodes different of the last level, two bits are needed. One bit indicates terminal node, and the other the inter/intraframe decision. For instance, if a 0 represents a terminal node, one alternative would be:
00: terminal interframe node.    01: terminal intraframe node.

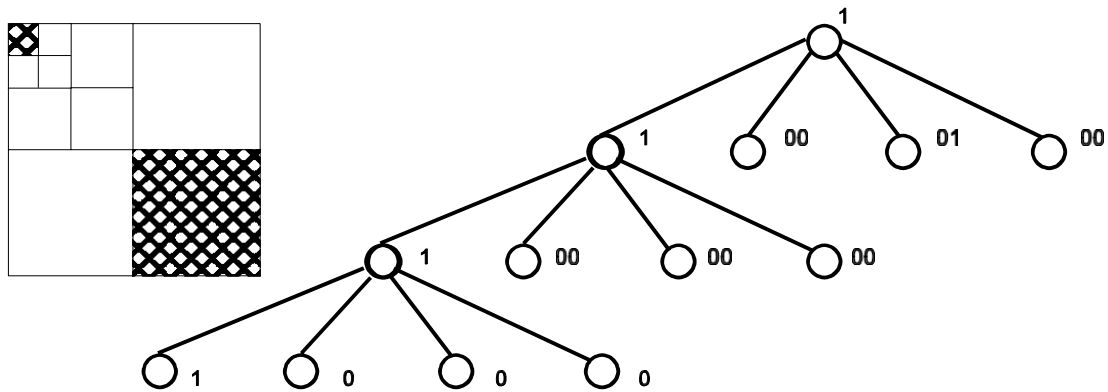

fig. 2 modified quadtree for inter/intraframe predictor. Last level and second bit of terminal nodes codifies the inter/intraframe decision. Dashed squares are intraframe coded zones.

Table 1 summarizes the theoretical compressions obtained with the proposed algorithms for the interframe and the inter/intraframe cases.

| CASES | INTERFRAME (N=128) | | INTER/INTRAFRAME | |
|---|---|---|---|---|
| | WITHOUT FLAG | WITH FLAG | WITHOUT FLAG | WITH FLAG |
| BETTER | 100:7 (N:9) | 100:8 (N:10) | 100:12 (16:2) | 100:19 (16:3) |
| WORST | 100:104 (N:133) | 100:86 (N:110) | 100:131 (16:21) | 100:94 (16:15) |

table 1

For the case of inter/intraframe predictor we have only considered the reduction in the overhead

___________________________________________________________________________

information (bit for inter/intra decision), because the total compression depends on the intraframe predictor used.

Worst case situation is evaluated when it is still useful the modified quadtree structure. A flag is also used to indicate if the quadtree codification is selected or not (in order to prevent the cases when the quadtree does not imply a reduction).

For the inter/intraframe predictor, the error obtained with the intraframe predictor can be penalized (P) to increase the compression ratio, because the compression is bigger with the interframe method (0.03 bits per pixel in the case of block size 16x16 pixels). This penalization would be adaptive as function of the contiguous blocks, increasing it when the cluster is dependent of the considered inter/intra decisor.

condition of selection: $error_{interframe} < P*error_{intraframe}$ (P>1)

V. RESULTS

A sample of the obtained results is the following:

For a 16x16 pels block size, the results obtained for the two first images of the secretary sequence are summarized in table 2.

The number of bits needed for quadtree coding is:

$$16(roots) + 8 \cdot 4 (1^{st} leaves) = 48\ bits$$

which added to the number of bits needed for coding the vector information results:

$$\frac{48}{8} + 8 \cdot 15 + 68 = 97\ bytes$$

instead of the 256 bytes required originally. This implies a reduction of 100:37.9

| BLOCK SIZE | Nº BLOCKS | EQ. SUBIMAGES |
|---|---|---|
| 16x16 | 68 | 68 |
| 32X32 | 15 | 60 |
| 64X64 | 8 | 128 |
|  |  | 256 |

Table 2.

From table 2 it is deduced that in this case half of the image is coded with a block size of 64x64 (8 blocks which represent 128 blocks of 16x16 size).

If the motion estimation is realized with blocks of 8x8 on the same images, 268 bits are needed for the tree coding, and 514 bytes for the vectors. Thus, the reduction is 100:53 (table 3).

| BLOCK SIZE | Nº BLOCKS | EQ. SUBIMAGES |
|---|---|---|
| 8x8 | 412 | 412 |
| 16x16 | 89 | 356 |
| 32X32 | 12 | 192 |
| 64X64 | 1 | 64 |
|  |  | 1024 |

table 3.

Lower block size can be reduced to 4x4, but the modelation of the movement is not considerably better to the case 8x8 [5], and it is more difficult to group vectors, so it is not recommended to reduce the block size more than 8x8.

VI. OTHER POSSIBILITIES

These algorithms can be implemented with other tree structures, that let more flexibility in the block shape (for instance, rectangles), but more flexibility implies more bits for tree coding.

Other compression methods based on blocks, can utilize this structure to reduce the interblock redundancy, with a small adaptation. Interesting results have been obtained in the case of the block truncation coding [6], where the quadtree structure has been applied to postprocessing the information that must be encoded.

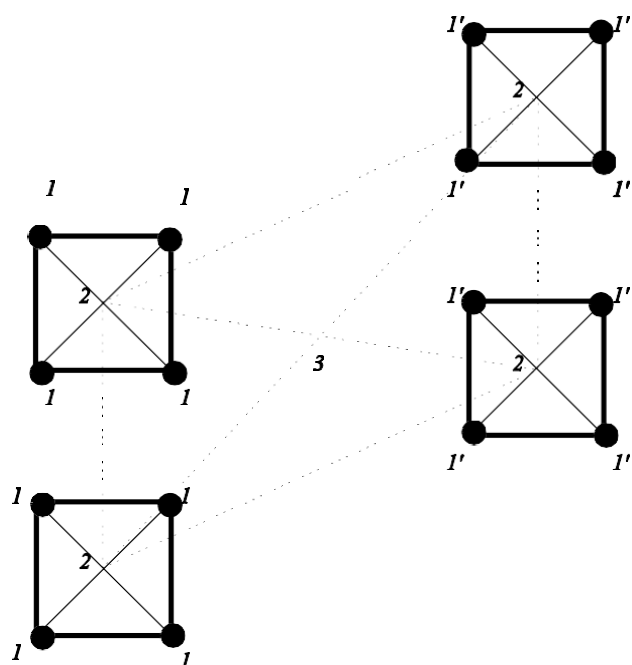

fig. 3  3D quadtree

___

The quadtree can be implemented in a 3D approach (see fig. 3), since vectors at the same location but different images ($I_k$ and $I_{k-1}$) are highly correlated. This assumption is correct if the objects have a constant motion along several frames. In this case, in the first stage 16 vectors are clustered into 4 vectors (vectors 1 of image $I_{k-1}$, and 1' of $I_k$). In the next step, the four vectors (referred as 2) are clustered into one vector (named 3).

Tree construction is similar to the process explained in section III.1, although the vectors belong to different images. Other tree structures or clustering strategies can be applied in a similar way. This method can also be applied to other coding methods that work with image sequences.